\documentclass{article}

\usepackage{arxiv}

\usepackage[utf8]{inputenc} 
\usepackage[T1]{fontenc}    
\usepackage{hyperref}       
\usepackage{url}            
\usepackage{booktabs}  
\usepackage{amsmath}
\usepackage{amsfonts}       
\usepackage{nicefrac}       
\usepackage{microtype}      
\usepackage{lipsum}
\usepackage{graphicx}
\usepackage{color}
\graphicspath{ {./images/} }
\usepackage[ruled,vlined]{algorithm2e}
\newcommand{\tabref}[1]{{Table \ref{#1}}}
\newcommand{\secref}[1]{{Section \ref{#1}}}

\newcommand{\figref}[1]{{Figure \ref{#1}}}

\newcommand{\DS}{{USM-SED}}

\newcommand{\FSD}{{FSD50K}}

\title{\DS - A Dataset for Polyphonic Sound Event Detection in Urban Sound Monitoring Scenarios}

\author{
Jakob Abeßer \\
  Semantic Music Technologies Group\\
  Fraunhofer IDMT \\
  Ilmenau, Germany \\
  \texttt{jakob.abesser@idmt.fraunhofer.de} 
 
}

\begin{document}
\maketitle
\begin{abstract}
This paper introduces a novel dataset for polyphonic sound event detection in urban sound monitoring use-cases. Based on isolated sounds taken from the \FSD~dataset, 20,000 polyphonic soundscapes are synthesized with sounds being randomly positioned in the stereo panorama using different loudness levels.
The paper gives a detailed discussion of possible application scenarios, explains the dataset generation process in detail, and discusses current limitations of the proposed \DS~dataset.
\end{abstract}


\section{Introduction}

Sound event detection (SED) is a crucial step in the analysis of polyphonic soundscapes, which surround us every day~\cite{Virtanen:2018:SoundSceneBook:BOOK}. 
The human auditory system can easily identify and focus on particular sounds sources around us.
In contrast, computational SED methods still struggle to recognize sounds in polyphonic mixtures \cite{Xia:2019:SEDSurvey:CSSP} for several reasons.
First, environmental sounds exhibit a large range of different timbre characteristics such as short transients, noise, and harmonic signal components.
Sound events of interest range from nature sounds like bird calls, rustling leaves, and rain drops, over machine-made sounds such as motor engines, braking noises, or chainsaws, to human-made sounds like voice, laughter, or screams. Such sounds range from structured or unstructured sounds, which can be stationary or non-stationary, repetitive or without any predictable nature. 
Second, the duration of acoustic events cover a large range from very short (gun shot, door knock, or shouts) to very long and almost stationary (running machines or climate sounds such as wind or rain). Since most SED methods analyze fixed sized audio segments, the classification of acoustic events is complicated, if the sound duration exceeds the analyzed time window.
Third, the temporal boundaries of environmental sounds are often ill-defined, which complicates their precise annotation and detection. 
For example, the loudness of passing vehicles like cars or trains gradually increases in the beginning and continuously fades into the background noise level at the end.
Fourth, sound events appear either in the foreground or background depending on the relative position of the corresponding sound sources within an acoustic scene.
If multiple sounds appear simultaneously, they overlap and often blend into novel mixture sounds, which complicates the sound classification.

Especially in urban environments, various sounds originating from traffic, vehicles, construction sites, and alarms cause an increase in the perceived amount of noise. 
Such a permanent noise exposure can have serious health implications for city residents. 
Over the last years, automatic noise monitoring systems were developed as countermeasures to identify the most salient sound sources, which contribute most to the overall noise perception \cite{Bello:2018:SONYC:CACM, Abesser:2019:Stadtlaerm:DCASE}.

As the main contribution of this paper, the \DS~dataset is presented, which is a novel dataset of short 5-seconds long polyphonic soundscapes. These soundscapes were created by systematically mixing isolated sound recordings taken from the public FSD50K dataset \cite{Fonseca:2020:FSD50K:ARXIV}.
In particular, a subset of 27 sound classes was defined with particular focus on urban soundscapes.
The \DS~dataset is intended as public benchmark for various machine listening tasks such as sound event detection and localization, source separation, and sound polyphony estimation as will be further detailed in \secref{ref:use_cases}. In order to foster reproducible research, we publish the Python scripts to synthesize the soundscapes of the \DS~dataset from the FSD50K dataset\footnote{\url{https://github.com/jakobabesser/USM_SED/}}.

\section{Urban Sound Monitoring Tasks}
\label{ref:use_cases}

In this section, several research tasks will be discussed, for which the \DS~dataset poses a novel and challenging evaluation dataset.

\subsection{Sound Event Detection} 
\label{sec:sed}

Several application scenarios in urban environments require the detection of sound events. In traffic monitoring, the types of passing vehicles such as cars, trucks, and busses need to be distinguished. Public events can be monitored to recognize sound events such as gunshots or bomb explosion to anticipate panic situations.
Construction site monitoring allows to synchronize the real construction progress with existing construction schedules.
Security monitoring applications require to detect sound events, which can indicate burglary into buildings or apartments such as broken glass of destroyed windows.
Similarly, SED algorithms also help to recognize different animal species for ecological monitoring tasks.

By focusing on polyphonic soundscapes, the \DS~dataset poses a big challenge to current SED algorithms and allows to systematically evaluate their performance depending on the soundscape polyphony level.
For evaluation purpose, both traditional SED metrics such as F-score or error rate \cite{Mesaros:2016:SEDMetrics:AS} as well as recently proposed metrics such as the polyphonic sound detection score (PSDS) \cite{Bilen:2020:AEDEval:ICASSP} are applicable.
One interesting research question, which can be studied using the \DS~dataset, is how training with monophonic samples (sound stems) compared to training with polyphonic mixtures influences the system's performance for polyphonic SED. Another interesting question is how the SED performance for a particular sound class is affected by it's dominance in the soundscape, i.\,e., whether the sound appears in the foreground or background.

\subsection{Source Counting / Polyphony Estimation}  
A second task, which is relevant to better understand polyphonic soundscapes, is to measure the number of active sound sources, i.\,e., the sound polyphony at a certain point in time.
A common definition of ``polyphony'' in the field of music analysis is the number of simultaneously sounding note pitches. In music information retrieval (MIR), this task was recently approached by combining musically-informed signal representations such as the Constant-Q transform (CQT) with different variants of convolutional neural networks (CNN) \cite{Taenzer:2021:LocalPolyphony:ELECTRONICS}.

As will be described in \secref{sec:dataset_creation}, the \DS~dataset includes 5 seconds long soundscapes with weak label annotations. 
Therefore, the definition of ``sound polyphony'' can be relaxed in such way, that it measures the number of active sounds within a short (5 s) duration instead of the number of simultaneously active sounds. 

\begin{table}[t]
 \caption{Comparison between the FSD50K and the \DS~datasets.}
  \centering
  \begin{tabular}{lll}
    \toprule
         & \textbf{\FSD}    & \textbf{\DS} \\
    \midrule
    Sound duration & 0.3 s - 30 s (variable) & 5 s (fixed)\\
    Labeling & weak & weak \\
    \# sound classes & 200 & 27 \\
    \# audio files & 51,197 & 20,000 \\ 
    Polyphony degree & 1 (monophonic) & 2-6 (polyphonic) \\
    Dataset size & 31.2 GB (sounds) & 49.3 GB (soundscapes + sounds) \\
    
    \bottomrule
  \end{tabular}
  \label{tab:comparison}
\end{table}

\subsection{Sound Event Localization}  

The task of sound event localization aims for a spatial localization of different sound sources, i.\,e., measuring their azimuth and elevation, relative to the audio recording location.
Datasets such as the one used in the DCASE 2019 challenge task ``Sound Event Localization and Detection''\footnote{\url{http://dcase.community/challenge2019/task-sound-event-localization-and-detection}} combine room impulse responses (RIR) measured in different recording locations, ambient (non-directional) noise components, and a synthetic mixing of polyphonic soundscapes \cite{Politis:2021:SEL:TASLP}.
In the \DS~dataset, a simpler approach was followed for soundscape synthesis.
Sound sources were positioned in the stereo panorama using the level difference approach as will be explained in \secref{sec:dataset_creation}.
Hence, the dataset can be used to estimate the stereo position of sound sources after their detection.

\subsection{Source Separation} 
Previous research on audio source separation has been mainly focused on speech signals \cite{Wang:2018:SpeechSeparation:TASLP} and music signals \cite{Cano:2019:SourceSep:SPM}.
Only recently, researches began to investigate the application of source separation algorithms on environmental sound mixtures \cite{Sudo:2019:EnvSoundSeg:IROS}.
Since the \DS~dataset includes both the audio mix (soundscape) and the corresponding stems, it provides a suitable test-bed for source separation algorithms in order to further stimulate research in this direction.
As one interesting application for noise monitoring applications, the individual noise level contributions of isolated sounds in a polyphonic soundscape could be measured after separating them from the mixture. This could be later used to weight the perceptual annoyance of individual sounds in complex soundscapes.

\begin{table}[t]
 \caption{Mapping between \FSD~sound classes (third column) to 27 sound classes included in the \DS~dataset (second column), which are grouped to six sound categories (first column).}
  \centering
  \begin{tabular}{p{.25\textwidth}p{.22\textwidth}p{.45\textwidth}}
    \toprule
    \textbf{Sound Category (\# classes)} & \textbf{\DS} & \textbf{\FSD} \\
         & \textbf{Class}     & \textbf{Classes} \\
    \midrule
    Rare sound events (7) & siren & ambulance (siren), emergency vehicle, fire truck, siren \\
    & gunshot & gunfire, machine gun \\
    & glass break & glass, shatter \\
    & church bell & church bell \\
    & alarm & Alarm, car alarm \\
    & lawn mower & lawn mower \\
    & spray can & spray \\
    \midrule
    Climate sounds (3) & wind & wind \\
    & rain & rain \\
    & thunderstorm & thunder, thunderstorm \\
    \midrule
    Animal sounds (2) & birds & bird \\
    & dogs & bark \\
    \midrule
    Human-made (4) & music & music \\
    & singing/cheering/applause & applause, booing, cheering, crowd \\
   & speech & kid speaking, conversation, woman speaking, male speech, man speaking, speech \\
   & scream & screaming, shout \\
    \midrule
   Construction site (4) & sawing & chainsaw, sawing \\
   & hammer & hammer \\
   & jackhammer & jackhammer \\
   & drilling & drill, power tool \\
    \midrule
   Vehicles (7) & car & car \\
    & truck & truck \\
   & bus & bus \\
   & motorcycle & motorcycle \\
   & train/tram & underground, train \\
   & airplane & aircraft engine, airplane \\
   & helicopter & helicopter \\
    \bottomrule
  \end{tabular}
  \label{tab:class_taxonomy}
\end{table}

\section{\DS~Dataset}

\tabref{tab:comparison} provides a general comparison between the \DS~and the \FSD~dataset.
While the \FSD~dataset includes monophonic sound recordings from a large number of sound classes (200), the \DS~dataset focuses polyphonic soundscapes which cover a smaller number of sound classes (27).
In this section, the \DS~dataset will be introduced in detail. 
First, \secref{sec:class_tax} discusses the applied sound class taxonomy.
Then, \secref{sec:dataset_creation} presents the algorithm, which was used to synthesize polyphonic soundscapes based on isolated sounds taken from the \FSD~dataset.
Finally, \secref{sec:critical_discussion} provides a critical discussion about current limitations of the \DS~dataset.

\subsection{Class Taxononmy}
\label{sec:class_tax}

As shown in \tabref{tab:class_taxonomy}, 27 sound classes were defined with relevance to urban sound monitoring research.
For each of these sound classes, mappings to one or multiple semantically corresponding sound classes in the \FSD~dataset were identified.
These 27 sound classes can be grouped into six categories: rare sound events, climate sounds, animal sounds, human-made sounds, construction site sounds, and vehicle sounds.
SED algorithms that are trained on these sound classes can be applied in the various acoustic monitoring scenarios discussed in \secref{sec:sed}: traffic monitoring, construction site monitoring, security monitoring, and ecological monitoring.

\subsection{Soundscape Rendering}
\label{sec:dataset_creation}

\begin{figure}[h!]
\centering
\includegraphics[width=.8\linewidth]{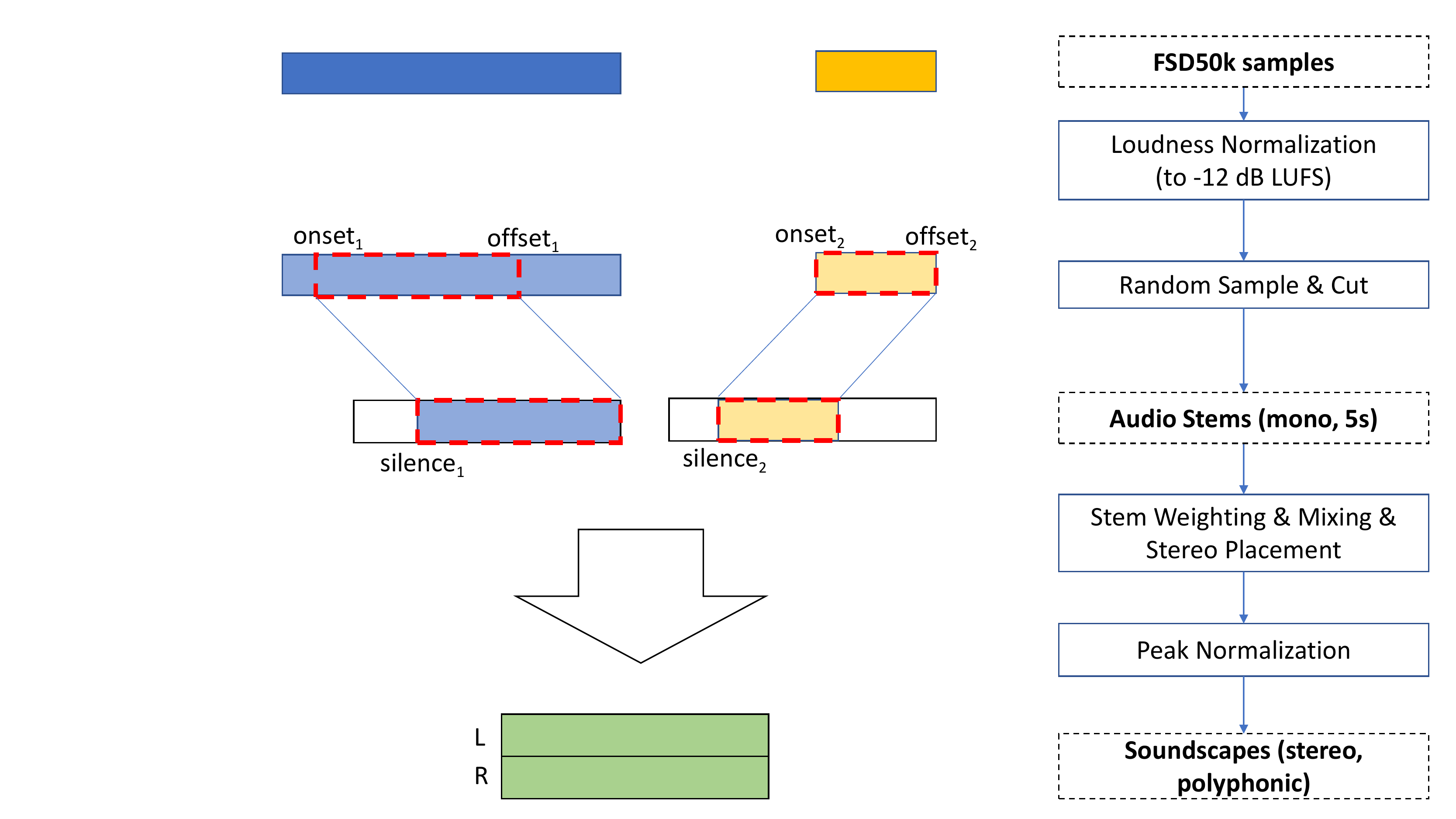}
\label{fig:rendering}
\caption{Audio synthesis steps to render polyphonic soundscapes as described in \secref{sec:dataset_creation}. Sample cropping procedure is illustrated for an audio sample longer than 5 s (blue) and an audio sample shorter than 5 s (yellow). }
\end{figure}

\begin{figure}[t]
\centering
\includegraphics[width=.7\linewidth]{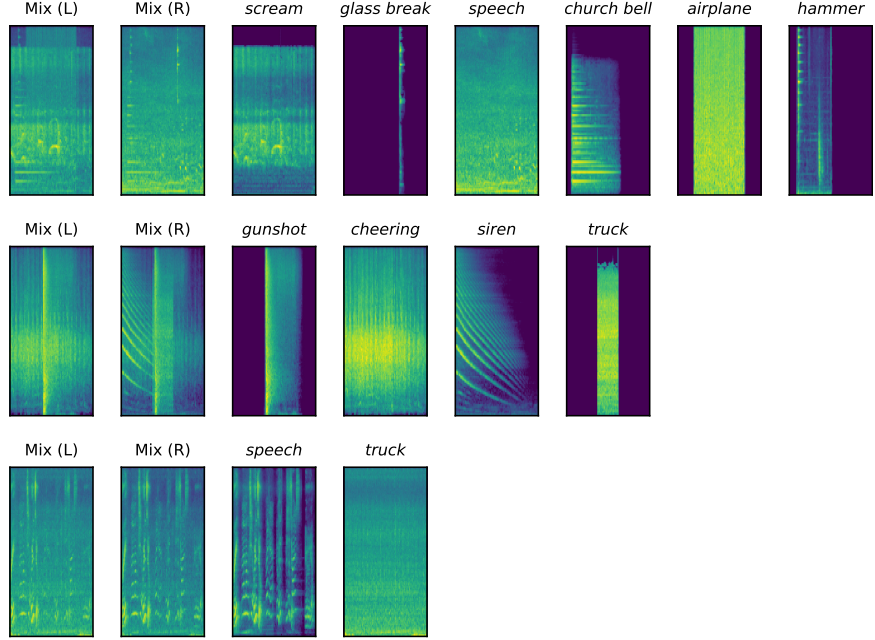}
 \label{fig:flowchart}
  \caption{Log-magnitude mel spectrograms (extracted using a hop size of 1024, windows size of 2048, and 128 mel bands) for stereo soundscapes with IDs 2417, 1930, and 339 taken from the \DS~ evaluation set. The examples are sorted according to their polyphony level (6, 4, and 2) in descending order.
  For each soundscape, the first two columns show the left and right channel and the additional columns show the included sounds (stems). }
  \end{figure}

\figref{fig:rendering} outlines the iterative procedure, which is used to synthesize 20,000 polyphonic stereo soundscapes of 
$5$ seconds duration, which are included in the \DS~dataset. A detailed description is provided in the remainder of this section.
The development subset is derived from samples of the \FSD~development set and includes a training set 
of 8,000 soundscapes and a validation set 
of 2,000 soundscapes.
The the evaluation subset 
of 10,000 samples is derived from the \FSD~evaluation set. For synthesizing the $i$-th soundscape, we proceed as follows:

\begin{enumerate}
    \item Randomly set the number of  sounds $N_F^i \in \left[1:3\right]$ mixed in the foreground and sounds $N_B^i \in \left[1:3\right]$ mixed in the background. The resulting level of polyphony is $L^i = N_F^i + N_B^i$.
    
    \item Randomly sample $L^i$ source samples each from a unique \DS~sound class and use $N_F$ of them them as foreground sounds and the remaining ones as background sounds. For the development set, we only sample from the development set of the \FSD~set, and vice versa for the evaluation set.
    
    \item Randomly crop a 5 s stem from each selected source sample. If the original sample duration is smaller than 5 s, place it at random position within the 5 s (this may cause a silence part in the beginning). If the original sample duration is larger than 5 seconds, randomly crop a segment of 5 seconds from it. The new sample arrays of 5 seconds duration are denoted as $x^i_j$ with $j$ indexing the stems of the current soundscape.
    Both cases are illustrated in \figref{fig:flowchart}.
    Only source samples of the \FSD~dataset with a duration smaller than 15 seconds are considered here to reduce the risk of label noise created by the sample cropping procedure.
    
    \item Use the \textit{pyloudnorm} Python package to normalize all stems $x^{i,j}$ to the same perceived loudness of -12 dB LUFS based on ITU-R BS.1770-4 specification\footnote{\url{https://www.itu.int/dms_pubrec/itu-r/rec/bs/R-REC-BS.1770-4-201510-I!!PDF-E.pdf}}.
    
    \item Randomly sample mixing coefficients  $\alpha^{i,j}$ as $\alpha^{i,j} \in \left[ -20, -8 \right]$ dB for the background sounds and $\alpha^{i,j}$ as $\alpha^{i,j} \in \left[ -6, 0 \right]$ dB for the foreground sounds.
    
    \item Randomly sample stereo panning coefficients $\beta^{i,j} \in \left[0, 1\right]$ for each sound ($0$ indicating left panning and $1$ indicating right panning).

    \item Render polyphonic stereo soundscapes from stems at a sample rate of $f_s=44.1$ kHz.
    
    \begin{enumerate}
    
    \item Compute linear mixing coefficients $\hat \alpha^{i,j} = 10^\frac{\alpha^{i,j}}{20}$.
    
    \item Normalize mixing coefficients as $\hat\alpha^{i,j} \leftarrow \frac{\hat\alpha^{i,j}}{\sum_j \hat\alpha^{i,j}}$.
    
    \item Mix mono samples to stereo file as 
    $s^i_0 = \sum_{j=1}^{L^i} (1- \beta^{i,j}) \cdot \hat \alpha^{i,j} \cdot x^{i,j}$ (left channel) and \linebreak
    $s^i_1 = \sum_{j=1}^{L^i} \beta^{i,j} \cdot \hat \alpha^{i,j} \cdot x^{i,j}$ (right channel).
    
    \end{enumerate}

    \item Define multi-label target vector $y^i$ with ones indicating all sound classes associated to the soundscape stems.
    
\end{enumerate}

\subsection{Critical Discussion \& Dataset Limitations}
\label{sec:critical_discussion}
  
The dataset generation procedure explained in \secref{sec:dataset_creation} goes along with several disadvantages and limitations, which will be discussed in this section.

\paragraph{Fixed soundscape duration}

Since sound events have a wide range of durations, the choice of a fixed sample duration of 5 s (such as in the ESC-50 dataset \cite{Piczak:2015:ESC50:ACM}) might truncate longer sounds and make them harder to recognize. The choice of 5 s represents a trade-off between common sound durations (compare Fig. 5, \cite{Fonseca:2020:FSD50K:ARXIV}) and the requirement for near real-time sound recognition scenarios as discussed in \secref{ref:use_cases}, where sound recognition results need to be updated around every 1-2 seconds. 

\paragraph{Stereo sound source placement}

In contrast to real-life soundscapes, where sound sources such as vehicles are moving, sounds in the \FSD~dataset are located as static sources at random positions in the stereo panorama.
Also, restricting the dataset to a stereo setup with two audio channels is a clear simplification compared to similar datasets for sound event localization such as the ``TAU Spatial Sound Events 2019 - Ambisonic'' and ``TAU Spatial Sound Events 2019 - Microphone Array''\footnote{\url{http://dcase.community/challenge2019/task-sound-event-localization-and-detection}}, which usually include spatial audio recordings with multiple audio channels.
Similarly to the fixed soundscape duration of 5 seconds, the choice of a stereo setup is motivated by practical considerations of low-cost acoustic sensors in urban sound monitoring scenarios.

\paragraph{Noise \& Label Noise \& Microphone Characteristics}

As consequence of the mixing process, the \DS~soundscapes directly derive from the audio samples in the \FSD~dataset.
The loudness normalization of these samples prior to the soundscape mixing can potentially boost the underlying noise levels.
Also, existing label noise based on incomplete or erroneous annotations directly propagates to the \DS~dataset.
Audio samples in the \FSD~dataset come from different uploaders in the FreeSound database and hence are recorded with different microphone setups \cite{Fonseca:2020:FSD50K:ARXIV}. The mixing procedure explained in \secref{sec:dataset_creation} consequently can lead to unrealistic blendings of different microphones characteristics. 
Nevertheless, a positive side-effect might be that this allows to train SED models, which are more robust to changes in recording conditions.

\paragraph{Licences}

All original sound samples from the \FSD~dataset, which were used in the creation of the \DS~dataset were published with one of the four licences ``CC BY'', ``CC BY-NC'', ``CC Zero'', or ``CC sampling+'', which all allow to remix and modify the original content. The USM-SED dataset is published alongside with a detailed list of all original FSD50k samples and their respective licences.

\paragraph{Soundscape Realism}

While the random selection of source samples in the mixing procedure often leads to unrealistic soundscapes, the \DS~dataset allows to train SED models, which are unbiased towards common sound co-occurrances in real-life scenarios.
A systematic investigation of common sound co-occurrences in different recording location remains an open question for machine listening research.

\paragraph{Scaper}

Salamon et al. published in \cite{Salamon:2017:Scaper:WASPAA} the Scaper library for soundscape synthesis, which was adapted in recent works on SED dataset generation \cite{Johnson:2021:FL:ARXIV}.
It covers most requirements, which arise from the dataset creation process described in \secref{sec:dataset_creation}. However, there are two main differences in both synthesis approaches.
First, in Scaper, a continuous texture-like sound is required as background sound\footnote{\url{https://scaper.readthedocs.io/en/latest/tutorial.html\#organize-your-audio-files-source-material}}, which should not contain any salient sound events and is therefore not included in the sound annotation of the resulting soundscape.
Since once goal of the \DS~dataset is to enable SED for both  foreground and background sounds, we aim for a complete annotation of all audible sound events allowing the \DS~dataset to be used for polyphonic SED evaluation.
Secondly, the Scaper library was not designed to randomly position individual sound sources in the stereo panorama.






\section*{Acknowledgements}

Graditude is due to Eduardo Fonseca, Xavier Favory, Jordi Pons, Frederic Font, and Xavier Serra for publishing the FSD50K dataset to the public.

\bibliographystyle{unsrt}  
\bibliography{refs_smt, refs_misc}  


\end{document}